# A rigorous benchmarking of methods for SARS-CoV-2 lineage abundance estimation in wastewater

**Authors and affiliations (unordered list)**

Viorel Munteanu
Department of Computers, Informatics, and Microelectronics, Technical University of Moldova, Chisinau, 2045, Republic of Moldova
viorelmunteanu.md@gmail.com
ORCID: https://orcid.org/0000-0002-4133-5945

Victor Gordeev
Department of Computers, Informatics, and Microelectronics, Technical University of Moldova, Chisinau, 2045, Republic of Moldova
victorgordeev.2010@gmail.com
ORCID: https://orcid.org/0009-0005-1052-2552

Michael Saldana
Astani Department of Civil and Environmental Engineering, University of Southern California, 3620 South Vermont Avenue, Los Angeles, CA 90089, USA.
michael.saldana.0@usc.edu
ORCID: https://orcid.org/0009-0007-7253-9434

Eva Aßmann
Genome Competence Center (MF1), Method Development and Research Infrastructure, Robert Koch Institute, 13353 Berlin, Germany
Centre for Artificial Intelligence in Public Health Research (ZKI-PH), Robert Koch Institute, 13353 Berlin, Germany
AssmannE@rki.de
ORCID: https://orcid.org/0000-0002-7249-069X

Justin Maine Su
Titus Family Department of Clinical Pharmacy, USC Alfred E. Mann School of Pharmacy and Pharmaceutical Sciences, University of Southern California, Los Angeles, CA 90089, USA
jmsu@usc.edu
ORCID: https://orcid.org/0000-0002-0912-4785

Nicolae Drabcinski
Department of Computers, Informatics, and Microelectronics, Technical University of Moldova, Chisinau, 2045, Republic of Moldova
drabcinski@gmail.com
ORCID: https://orcid.org/0009-0008-4381-836X

Oksana Zlenko
National Scientific Center "Institute of Experimental and Clinical Veterinary Medicine", Kharkiv, 61023, Ukraine
oksana.ceratium@gmail.com
ORCID: https://orcid.org/0000-0002-4357-0762

Maryna Kit
National Scientific Center "Institute of Experimental and Clinical Veterinary Medicine", Kharkiv, 61023, Ukraine
maryna_kit@ukr.net
ORCID: https://orcid.org/0000-0003-4306-7611

Khooshbu Kantibhai Patel
Titus Family Department of Clinical Pharmacy, USC Alfred E. Mann School of Pharmacy and Pharmaceutical Sciences, University of Southern California, Los Angeles, CA 90089, USA
khooshbu@usc.edu
ORCID: https://orcid.org/0000-0002-7575-4734

Yidian Xu
Department of Translational Genomics, Keck School of Medicine, University of Southern California, 1450 Biggy Street, Los Angeles, CA, USA
XYD458@gmail.com
ORCID iD: https://orcid.org/0009-0003-1340-360X

Abdullah Al Nahid
Department of Biochemistry and Molecular Biology, School of Life Sciences, Shahjalal University of Science and Technology, Sylhet 3114, Bangladesh
abdnahid56@gmail.com
ORCID: https://orcid.org/0000-0002-4390-0768

Pavel Skums
Department of Computer Science and Engineering, University of Connecticut, Storrs, Connecticut
pavel.skums@uconn.edu
ORCID: https://orcid.org/0000-0003-4007-5624

Shelesh Agrawal
Chair of Water and Environmental Biotechnology, Institute IWAR
Department of Civil and Environmental Engineering Sciences, Technical University of Darmstadt, Darmstadt, Germany
s.agrawal@iwar.tu-darmstadt.de
ORCID: https://orcid.org/0000-0001-9365-5951

Martin Hölzer
Genome Competence Center (MF1), Method Development and Research Infrastructure, Robert Koch Institute, 13353 Berlin, Germany
HoelzerM@rki.de
ORCID: https://orcid.org/0000-0001-7090-8717

Adam Smith
Astani Department of Civil and Environmental Engineering University of Southern California 920 Downey Way; BHE 221 Los Angeles, CA 90089
smithada@usc.edu
ORCID: https://orcid.org/0000-0002-3964-7544

Alex Zelikovsky
Department of Computer Science, College of Art and Science, Georgia State University, Atlanta, GA, USA
alex.zelikovsky@gmail.com
ORCID: https://orcid.org/0000-0003-4424-4691

Serghei Mangul*
Titus Family Department of Clinical Pharmacy, USC Alfred E. Mann School of Pharmacy and Pharmaceutical Sciences, University of Southern California, 1540 Alcazar Street, Los Angeles, CA 90033, USA
Department of Quantitative and Computational Biology, University of Southern California Dornsife College of Letters, Arts, and Sciences, Los Angeles, CA 90089, USA.
serghei.mangul@gmail.com
ORCID: https://orcid.org/0000-0003-4770-3443

*For correspondence: serghei.mangul@gmail.com

**Abstract**

In light of the continuous transmission and evolution of SARS-CoV-2 coupled with a significant decline in clinical testing, there is a pressing need for scalable, cost-effective, long-term, passive surveillance tools to effectively monitor viral variants circulating in the population. Wastewater genomic surveillance of SARS-CoV-2 has arrived as an alternative to clinical genomic surveillance, allowing to continuously monitor the prevalence of viral lineages in communities of various size at a fraction of the time, cost, and logistic effort and serving as an early warning system for emerging variants, critical for developed communities and especially for underserved ones. Importantly, lineage prevalence estimates obtained with this approach aren't distorted by biases related to clinical testing accessibility and participation. However, the relative performance of bioinformatics methods used to measure relative lineage abundances from wastewater sequencing data is unknown, preventing both the research community and public health authorities from making informed decisions regarding computational tool selection. Here, we perform comprehensive benchmarking of 18 bioinformatics methods for estimating the relative abundance of SARS-CoV-2 (sub)lineages in wastewater by using data from 36 *in vitro* mixtures of synthetic lineage and sublineage genomes. In addition, we use simulated data from 78 mixtures of lineages and sublineages co-occurring in the clinical setting with proportions mirroring their prevalence ratios observed in real data. Importantly, we investigate how the accuracy of the evaluated methods is impacted by the sequencing technology used, the associated error rate, the read length, read depth, but also by the exposure of the synthetic RNA mixtures to wastewater, with the goal of capturing the effects induced by the wastewater matrix, including RNA fragmentation and degradation.

**Introduction**

As the SARS-CoV-2 virus continues to spread and mutate, while the rates of COVID-19 clinical testing have plummeted,[1] there is a pressing need for scalable, cost-effective, sensitive, and long-term passive surveillance tools enabling continuous monitoring of the viral variants circulating in the population. Wastewater genomic surveillance (WWGS) of SARS-CoV-2 has demonstrated repeatedly its efficacy in tracking lineage prevalence dynamics in numerous countries worldwide,[2–6] including those with limited resources.[7,8] In contrast to clinical genomic surveillance, SARS-CoV-2 WWGS enables instant access to comprehensive and continuous data from population cross-sections of any size at a fraction of the cost. In addition, it captures asymptomatic cases and is free from biases associated with clinical testing accessibility and individual compliance or participation, therefore providing more accurate estimates of lineage prevalence within communities.[9–12] Furthermore, it promises to detect novel cryptic variants, including those resistant to naturally acquired or vaccine-induced immunity, those rarely observed in clinical samples, and those from unsampled infected individuals.[13,14] Most importantly, it enables detection of emerging viral variants earlier than clinical monitoring,[15–17] and consequently, WWGS can act as an early warning system of critical importance for public health officials and policymakers, providing them with advanced notice to prepare for potential outbreaks,[7,18] guiding the deployment of extensive response mechanisms, such as mass testing,[19] and ensuring effective resource allocation to manage and contain the spread of the virus. Given all these, WWGS promises to become an indispensable public health tool able to elucidate the geospatial distribution of viral lineages and outbreak clusters within communities as well as lineage prevalence, including prevalence time trends.

Despite the enormous potential of WWGS, the quality of the sequencing data obtained from wastewater samples poses significant challenges for its bioinformatics analysis, being impacted by the low viral loads, the heavy degradation and fragmentation of viral RNA,[15] a rich background of contaminating nucleic acids belonging to other species, as well as the highly variable chemistry and physics of wastewater, including the pH, temperature and mixing intensity. Moreover, different PCR-inhibitory compounds from wastewater networks[20–22] can interfere with enzymes from library preparation and amplification steps, which can lead to nonuniform sequencing depth and incomplete genome coverage. Additionally, sampling and laboratory methods can also impact the quality of sequencing data. These include the sampling strategy,[23–27] sample preservatives and storage conditions,[28,29] as well as the methods used for viral concentration,[30,31] RNA extraction,[31–33] and sequencing library preparation.[34] Finally, the quality of the sequencing data can be also affected by the choice of the sequencing technology, including the specific platform, read length, library preparation kit, and amplicon panels.[35] Importantly, it has not been explored enough how varying laboratory methods and commercially available kits but also the complex and variable properties of wastewater samples impact the accuracy and reliability of subsequent bioinformatics analysis.

In response to the pandemic, numerous methods were developed for relative abundance estimation of SARS-CoV-2 lineages. Some of these, including methods such as Freyja,[15] VaQuERo,[7] LCS,[36] PiGx,[37] and Alcov[38] are based on deconvolution algorithms, which are applied to a list of mutation frequencies obtained by aligning the reads to a reference. In a sample with several viral lineages, the expected fraction of mutated reads at any given locus is equal to the sum of the relative abundances of lineages carrying this particular mutation. A constrained regression method is then applied to estimate the relative contribution of a reference set of lineages to the observed distribution of mutation frequencies. Another group of methods which was developed for lineage quantification employs classification algorithms – these are methods such as COJAC[39] or the VLQ pipeline[40] (based on Kallisto[41]). It also includes methods using the Expectation-Maximization algorithm to obtain maximum likelihood estimates of the proportions of various haplotypes present in the sample[42]. These classification algorithms assign each read either deterministically or probabilistically to individual reference lineages according to the signature mutations they share and then aggregate the read counts by lineage to estimate their total abundance.

Some bioinformatics tools, such as Salmon[43] and Kallisto,[41] which were either used or considered for relative abundance estimation of SARS-CoV-2 lineages in wastewater, were originally designed for transcript quantification using RNA-Seq rather than for wastewater sequencing data analysis, and therefore, it remains largely unknown how well they can handle the genomic complexity and technical specifics characterizing the latter. While the authors of the VLQ pipeline[40] for SARS-CoV-2 lineage abundance estimation used Kallisto as their quantification engine, the accuracy of the latter was only briefly compared to other popular transcriptomics tools, namely Salmon,[43] RSEM,[44] and IsoEM2[45] by using simulated data. In another study,[46] the authors used simulated and empirical data to compare the performance of only five methods for SARS-CoV-2 lineage abundance estimation based on both classification and deconvolution approaches, which only covers a handful of the existing tools and pipelines used for this purpose.

In this study, we perform a comprehensive benchmark for measuring the accuracy of bioinformatics methods used for relative abundance estimation of SARS-CoV-2 (sub)lineages in wastewater samples. In particular, we will benchmark 18 bioinformatics methods, which include methods specifically designed for wastewater sequencing data analysis as well as transcriptomics tools repurposed for this task (Table 1).[41,43] We will systematically evaluate and compare the accuracy of these computational methods in estimating relative lineage abundances and how it is impacted by the design of the sequencing experiment, in particular by the sequencing technology used, the error rate, read length, and read depth, but also by different periods of exposure of the viral RNA to the wastewater matrix, and therefore also by the extent of RNA degradation and fragmentation. To achieve this, we will use a series of real and simulated datasets. In particular, we used synthetic RNA genomes to generate 16 *in vitro* mixtures containing combinations of SARS-CoV-2 lineages and sublineages according to standard ratios, which were sequenced directly using the Ion Torrent technology. Our simulated datasets are based on 78 complex *in silico* mixtures of lineages and sublineages co-occurring in the clinical setting with proportions mirroring the prevalence ratios observed in the real data. Importantly, a subset of these *in silico* mixtures will be used for *in vitro* generation of 20 real mixtures with roughly the same lineage composition and proportions by using synthetic RNA genomes. These mixtures will be sequenced directly but also incubated with wastewater for different periods followed by RNA extraction and Illumina sequencing. The results of this benchmark are meant to guide the selection of the most accurate and robust quantification tools that can be used for estimation of lineage prevalence in the population, including for early-on detection of known circulating and emerging lineages. We also hope that this collection of real and simulated wastewater sequencing datasets obtained for the combination of variables listed above will serve as an informative resource and a reference for the research community and the public health authorities.

**Methods**

**Generation of realistic *in silico* mixtures of lineages and sublineages**

To generate realistic mixtures that would mirror the prevalence ratios of viral (sub)lineages observed during the pandemic, we used weekly count data for co-occurring SARS-CoV-2 lineages and sublineages reported for the United States between January 2020 and December 2022 from the Global Initiative on Sharing All Influenza Data (GISAID) database.[47] Only (sub)lineages with a weekly case count greater than 10 were included in our analysis. A total of 11 co-occurring lineages were detected during this period: B.1.1.7, B.1.351, B.1.617.2, B.1.429, B.1.525, P.1, B.1.526, B.1.617.1, C.37, P.3, and P.2, as well as a total of 11 co-occurring omicron and delta sublineages and their parent lineages: AY.23, BQ.1, BA.4, BA.2.12.1, XBB.1.5, BA.2.75, AY.27, BA.2, XBB, BA.1, BA.5, and B.1.617.2. Reported weekly counts were normalized to percentages and the latter were used to generate a total of 42 time-ordered mixtures for the corresponding group of lineages and 36 for the group of delta and omicron sublineages and their parent lineages (Supplementary Tables 1 and 2, respectively). Each individual mixture represents the actual frequencies witnessed during a particular week of the above-mentioned period (Supplementary Fig. 1 and 2, respectively). A key characteristic of these mixtures is the high proportion of small relative abundance values (Supplementary Fig. 2a and b), with non-null values smaller than 1% representing around 47% of the values for the group of lineages (min=0.001%) and around 18% for the group of sublineages (min=0.007%).

**Generation of *in vitro* mixtures of lineages and sublineages**

To create realistic *in vitro* mixtures with roughly the same quantitative composition as our in silico samples, we picked subsets of 10 mixtures from the series of lineage and sublineage mixes described above. The approximate composition of these mixtures will be recreated *in vitro* by using synthetic lineage and sublineage genomes acquired from Twist Biosciences. Each of these synthetic genomes marketed as virus controls consists of six non-overlapping 5kb fragments of ssRNA, which are claimed to provide over 99.9% coverage of the SARS-CoV-2 genome. Given that not all of the naturally co-occurring (sub)lineages from our *in silico* mixtures had a synthetic genome equivalent in the offer of the respective company, a few adjustments had to be made. In particular, 4 lineages from the group of 11 naturally cooccurring lineages, namely B.1.525, C37, P.2, and P.3 were excluded, with their corresponding percentages being distributed to the remaining lineages in each mixture, while preserving the null values (Supplementary Table 4). Similarly, in the group of naturally cooccurring sublineages, the following 5 didn't have a synthetic equivalent: AY23, AY27, BA.2.75, XBB, and XBB.1.5. The latter three were excluded and their values were distributed to the remaining sublineages in each mixture, except to those with null abundances. At the same time, sublineages AY.23 and AY.27 were substituted by AY.1 and AY.2, respectively, with their corresponding relative abundances left unaltered (Supplementary Table 5). The genome references corresponding to the synthetic lineages and sublineages used to produce the mixtures are listed in Supplementary Table 6.

Next, the real mixtures will be pipetted using the synthetic RNA controls described above, which come in volumes of 100 µL each and at a concentration of approximately one million copies per microliter. These will be diluted in RN-ase free water and a corresponding amount of each control will be used for each mixture (Supplementary Tables 4 and 5). The obtained "clean" mixtures will be reverse transcribed and sequenced according to the procedure described below.

Furthermore, in order to assess the effects induced by the wastewater matrix on the performance of the benchmarked bioinformatics methods, known concentrations of the same "clean" mixtures will be added to a SARS-CoV-2-negative wastewater sample collected before the pandemic and stored in a frozen state. These mixtures will be incubated for different periods of time, followed by RNA extraction and sequencing, as described further below.

In addition to the realistic mixtures described above, a subset of the synthetic RNA genomes plus the wild type SARS-CoV-2 genome (Wuhan-Hu-1 A.1) were used to generate a series of 16 *in vitro* mixtures representing simple standard ratios of lineages and/or sublineages, where each (sub)lineage has a relative abundance of either 25% or 50% (Supplementary Table 7). These mixtures were not incubated with wastewater and were instead sequenced directly using the Ion Torrent technology as described below.

**Simulating reads from the *in silico* viral mixtures**

The reference genomes of the viral lineages and sublineages required for simulating the sequencing reads were downloaded from GISAID (Supplementary Table 3). These reference sequences and the mixture compositions described earlier will be used to simulate sequencing reads according to three strategies in order of increasing complexity of the data: (a) error-free reads (dataset 1); (b) reads with sequencing platform-specific errors (dataset 2), and (c) reads with sequencing errors that also mimic RNA fragmentation and degradation, in order to account for wastewater sample specifics (dataset 3). To ensure a comprehensive benchmarking analysis, the reads with sequencing errors and simulated RNA degradation (dataset 3) will be used in combination with different read depth and read length.

Illumina reads will be generated using the short-read simulation tool ART v2.5.8,[48] which uses empirical read quality profiles summarized from large real sequencing data. The list of ART parameters relevant for our study and their interpretation is provided in Supplementary Table 8. To generate paired-end sequencing data with read length of 150 bp, mean library fragment size of 300 bp (sd=10bp), and quality profiles corresponding to the HiSeq 2500 sequencing system (dataset 2), the following command will be used: *art_illumina -ss HS25 -ef -sam -i /path/to/genome/sequence/fasta/file -p -c nr_of_reads_to_generate -l 150 -m 300 -s 10 -ir 0.0001 -o output/path.* The inclusion of the -ef parameter in the ART command above will generate reads with ideal quality scores and no sequencing errors from identical genomic positions (dataset 1). The parameter -c will be used to indicate the number of read counts to be generated for each lineage and sublineage of each sample according to the ground truth relative abundance values listed in Supplementary Tables 1 and 2 and the total read count per sample. The latter was calculated to ensure a total read count of over one million reads.

The ART command above will be run to produce a SAM file containing a true alignment for each lineage present in the sample. SAM files will be merged with samtools v1.13[49] and the reads will be extracted to FASTQ files using Picard v3.1.0[50]. The reads in the FASTQ files will be aligned to the wild-type SARS-CoV-2 reference with the RefSeq accession NC_045512.2 (GenBank accession MN908947.3) using the BWA[51] v0.7.17 aligner with default parameters. The resulting SAM files will be also converted to BAM files with samtools using default parameters. For dataset 1, FASTQ, SAM, and BAM files will be created similarly.

To simulate sequencing data with read lengths of 36, 50, 75, 100, 150, and 250 bp, the same ART command will be used, but with parameter -l set to the listed values and parameter -ss set to the sequencing platform ID corresponding to each read length, as indicated by the ART manual. To generate data with read depth values of 10, 100, 250, 1,000, the c parameter will be modified accordingly compared to the values used for datasets 1 to 3. Furthermore, to assess the impact of targeted versus whole-genome sequencing on lineage quantification accuracy, data with a read depth of 10,000 will be generated from the corresponding region of the SARS-CoV-2 genome.

**RNA extraction and sequencing of the *in vitro* mixtures**

The synthetic RNA mixtures containing standard lineage ratios (Supplementary Table 7) were reverse transcribed using SuperScript™ VILO™ Master Mix (Thermofisher Scientific). Library preparation was carried out by using the Ion AmpliSeq SARS-CoV-2 Research Panel (Thermofisher Scientific), which includes 237 primer pairs and results in amplicons of 125-275 bp in length, ensuring near-complete coverage of the SARS-CoV-2 genome. Eight libraries were multiplexed per sequencing run and sequenced on an Ion S5 sequencer using an Ion Torrent 530 chip (Thermofisher Scientific). Multiple sequencing runs were performed to ensure a minimum of one million mapped reads per sample. Both library preparation and sequencing were performed according to the manufacturer's instructions.

The *in vitro* samples with lineage composition similar to the corresponding *in silico* samples, represented by both the "clean" RNA mixtures and the RNA extracted from wastewater-incubated mixtures will be sequenced using the Illumina technology.

**Choice and installation of lineage quantification methods**

We reviewed the scientific literature, including both published journals and preprint studies, for bioinformatics tools and pipelines capable of estimating the abundances of SARS-Cov-2 (sub)lineages from wastewater sequencing data. This includes the four transcriptomics tools mentioned earlier: Kallisto, Salmon, RSEM, and IsoEM2. We also browsed the web for implemented computational tools that do not appear in publications. Importantly, our list comprises computational methods that take different types of input, including mapped or unmapped sequencing reads, VCF files, or lists of mutations and their corresponding frequencies. We specifically ignored variant callers and tools that yield a list of alleles and their corresponding frequencies as output since obtaining lineage relative abundances using the latter requires additional algorithmic steps. Our final list contains a total of 18 tools (Table 1). These will be installed by closely following the accompanying documentation and recommendations issued by the authors. Successful installation will be confirmed by running the embedded examples where available and inspecting the output.

**Preprocessing of sequencing data**

Sequencing data will be pre-processed as follows: primer trimming will be accomplished with cutadapt[52] (v4.2), Trimmomatic[53], or a similar program and read alignment to the wild type reference genome will be done with minimap2[54] (v2.24), BWA[51] (v0.17), BWA-MEM2[55] (v2.3), or similar, depending on the recommendations of the authors of the tool or pipeline. Where author recommendations are missing, we will use the same software in order to avoid introducing variation at this step. For tools requiring primer trimming, we will supply the appropriate *.fasta or *.bed files with primers. FastQC[56] (v0.12.1) will be used to examine the quality of the raw reads, and the alignments will be visually inspected using IGV[57] (v2.16). For simulated reads, preprocessing steps such as primer trimming are not necessary, since the reads are generated without primers or adapters.

**Running the quantification methods**

All tools and pipelines will be run with default parameters, with the exceptions listed below. In particular, given that our primary goal is measuring the quantification accuracy of SARS-CoV-2 (sub)lineages, the exact set of (sub)lineages present in our samples is assumed to be known. Consequently, for tools that allow this, we will restrict the set of lineages to be searched for and quantified to the lists used to create our *in silico* and *in vitro* datasets. Also, considering the high proportion of low relative abundance values in both our simulated and real sequencing data (Fig. 2), we will run the tools in maximum sensitivity mode, which means that the cutoff values of the parameters that can prevent the detection of the lowest lineage abundances in our samples will be adapted correspondingly. All the tools and pipelines assessed in this study are open-source and can be accessed via link.

The tools in our list which were originally developed for use with transcriptomics data and were repurposed here for viral lineage quantification, namely Kallisto, Salmon, RSEM, and IsoEM2, will be given as input the genome sequences of the predefined set of (sub)lineages present in the corresponding dataset in place of the sequences of transcripts normally expected by these programs. For tools and pipelines which require a genome reference sequence as input, the wild-type NC_045512.2 reference will be used. Sequencing data input will be provided either as raw FASTQ files or preprocessed BAM or SAM files, depending on the requirements of each tool. For tools that require BAM or SAM files as input, we will preprocess the raw FASTQ reads by trimming the corresponding primers and aligning the reads to the NC_045512.2 wild-type reference. Any special input files required by the tools will be provided as necessary. For those computational methods that take a different input, including VCF files or lists of mutations detected in a sample and their corresponding frequencies, these input files will be generated from the sequencing data using the same tool(s) and procedure to avoid introducing variation at this step.

For the bioinformatics methods assessed in this study that require user-sourced mutation constellation files or lists of mutations for each lineage or sublineage to be quantified, we will create these using the same rules according to which other constellations used by this tool were created.

**Evaluation of the accuracy and performance of the quantification methods**

For each tool and dataset, the custom output files containing the relative abundance estimates will be parsed into the standard csv format identical to the format of the tables with ground truth values (Supplementary Tables 1, 2, 4, and 5). Where needed, non-standard lineage quantification output, including transcripts per million or absolute counts of lineage-specific markers, will be converted to percentages. Lineages not called despite being present in our samples (false negatives) will be assigned null values. Absolute errors (L1) will be computed by subtracting the ground truth relative abundance value from each corresponding value of the resulting table, and the obtained difference will be divided by the corresponding ground truth value to obtain relative errors. Additionally, for each tool and dataset, the mean absolute error (MAE) will be computed as the mean of the moduli of all absolute errors. The difference in accuracy of the benchmarked tools will then be assessed by directly comparing the MAE values obtained from the same dataset as well as the distributions of absolute and relative errors. The quantification method with the lowest MAEs across most datasets tested will be considered the most accurate.

For any given tool, MAEs or per-sample L1 errors will be also compared across different datasets to assess the impact of different tested conditions on its accuracy. In particular, MAE values will be compared between the datasets: a) with and without sequencing errors; b) with and without simulated RNA degradation; c) with different read lengths; d) with different read depths; e) with and without exposure to wastewater.

**Estimating performance**

To assess the performance of the computational methods, we will record the CPU time and RAM usage metrics for each of the tools by running these on 10 samples of the *in vitro* dataset and computing averages across samples.

**Data availability**

All simulated sequencing reads and the raw data obtained from sequencing the *in vitro* mixtures will be uploaded to the Sequence Read Archive and the corresponding accession number will be included in the manuscript. All data required to produce the figures and perform the analysis described in this paper will be also made freely available.

**Code availability**

We commit to share all code used for simulating the sequencing datasets, all code and special input files used to run the computational methods benchmarked in this study, and all code used to analyze the data and to generate the figures. The code will be made public under the Massachusetts Institute of Technology (MIT) license.

abundances from wastewater sequencing data. 2023.06.02.543047 Preprint at https://doi.org/10.1101/2023.06.02.543047 (2023).

**Table 1 | The list of benchmarked bioinformatics methods for measuring lineage relative abundances**

| Bioinformatics tool or pipeline | Citation | Web page |
|---|---|---|
| Kallisto | Bray et al., 2016[41] | https://github.com/pachterlab/kallisto |
| VLQ (VLQ-nf) | Baaijens et al., 2021[40] (Aßmann et al., 2023[66]) | https://github.com/baymlab/wastewater_analysis (https://github.com/rki-mf1/VLQ-nf) |
| Salmon | Patro et al., 2017[43] | https://salmon.readthedocs.io |
| Alcov | Ellmen et al., 2021[38] | https://github.com/Ellmen/alcov |
| Freyja | Karthikeyan et al., 2022[15] | https://github.com/andersen-lab/Freyja/blob/main |
| Gromstole | NA | https://github.com/PoonLab/gromstole |
| LCS | Valieris et al., 2022[58] | https://github.com/rvalieris/LCS |
| Lineagespot | Pechlivanis et al., 2022[59] | https://github.com/BiodataAnalysisGroup/lineagespot |
| unnamed | Pipes et al., 2022[42] | https://github.com/lpipes/SARS_CoV_2_wastewater_surveillance |
| RSEM | Li et al., 2011[44] | https://github.com/deweylab/RSEM |
| IsoEM2 | Mandric et al., 2017[45] | https://github.com/mandricigor/isoem2 |
| VaQuERo | Amman et al., 2022[7] | https://github.com/fabou-uobaf/VaQuERo |
| PiGx | Schumann et al., 2022[37] | https://github.com/BIMSBbioinfo/pigx_sars-cov-2 |
| wastewaterSPAdes | Korobeynikov et al., 2022[60] | https://github.com/ablab/spades |
| CovMix | NA | https://github.com/chrisquince/covmix |
| V-pipe (Lollipop) | Posada-Céspedes et al., 2021[62] (Dreifuss et al., 2022[61]) | https://github.com/cbg-ethz/V-pipe (https://github.com/cbg-ethz/LolliPop) |
| Aquascope (C-WAP successor) | Ewels et al., 2020[63] | https://github.com/CDCgov/aquascope |
| VirPool | Gafurov et al., 2022[65] | https://github.com/fmfi-compbio/virpool |

**Supplementary information**

A

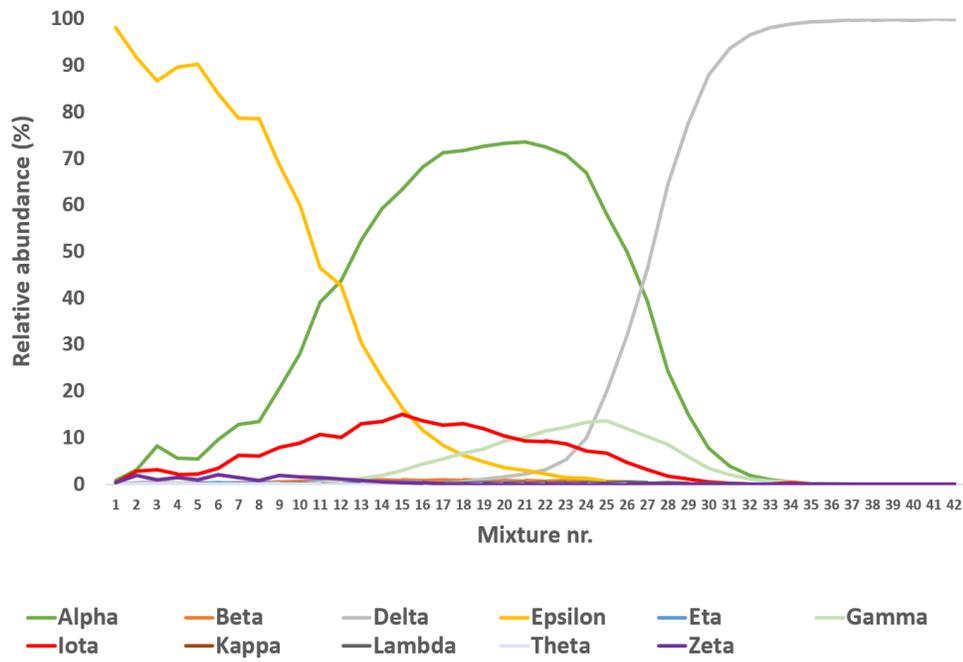

B

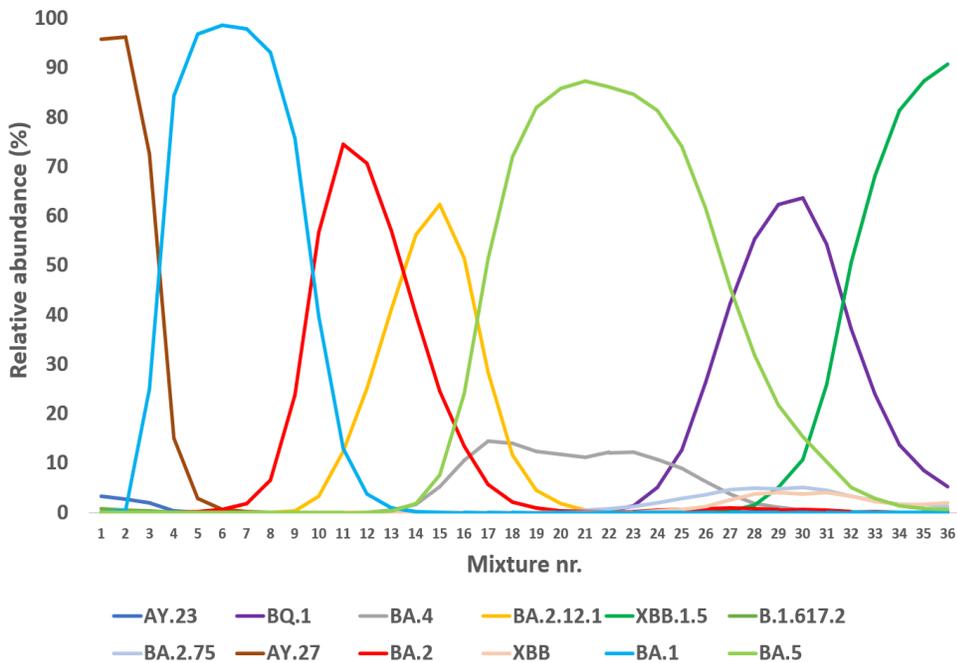

**Supplementary Fig. 1 | Line plots of the relative abundance values for the group of lineages (A) and the group of sublineages and their parent lineages (B) as seen in the time-ordered mixtures used to generate the simulated datasets.**

A

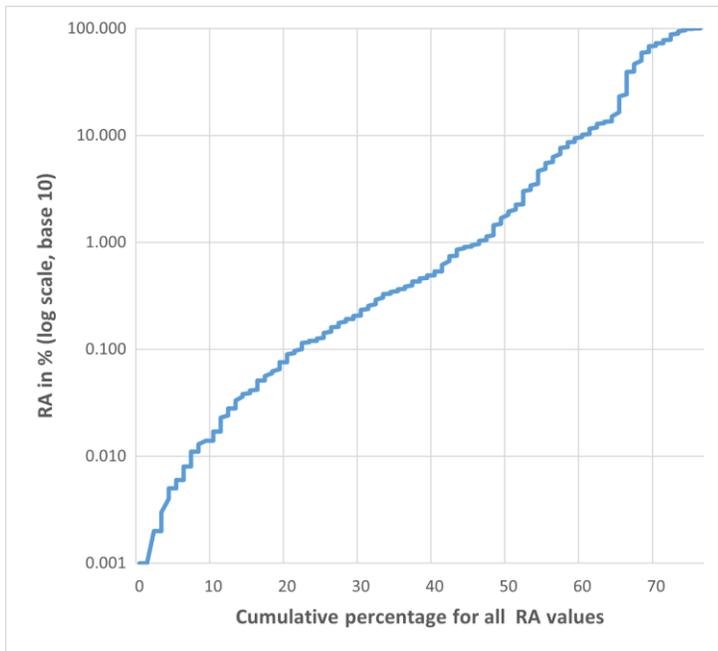

B

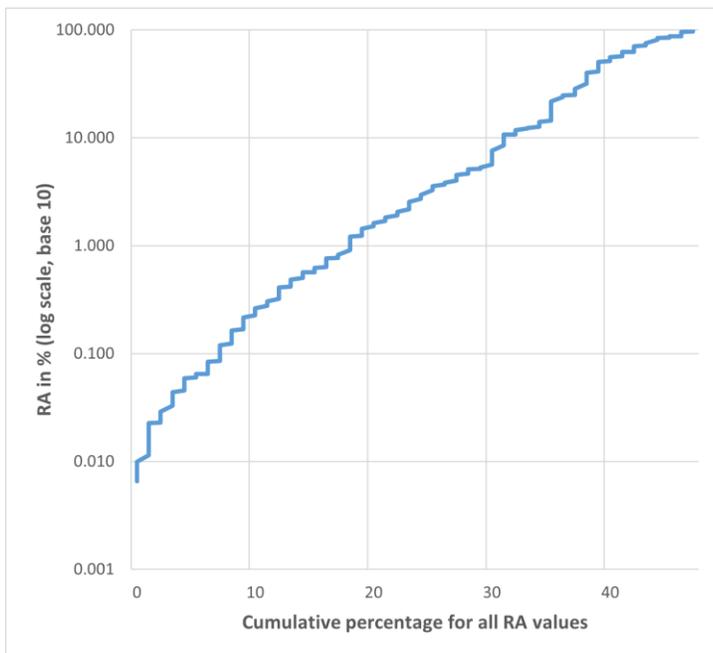

**Supplementary Fig. 2 | The distribution of the non-null relative abundance values in the mixtures of lineages (A) or the sublineages and their parent lineages (B) used to generate the simulated datasets.** Cumulative percentages are calculated based on all RA values, including null values, which represent around 24% and 52% of the total nr. of data points for lineages and sublineages, respectively.

**Supplementary Table 1 | The composition (in %) of the *in silico* mixtures of lineages**

|     | Alpha   | Beta    | Delta     | Epsilon | Eta     | Gamma   | Iota    | Kappa     | Lambda | Theta | Zeta  |
|-----|---------|---------|-----------|---------|---------|---------|---------|-----------|--------|-------|-------|
|     | B.1.1.7 | B.1.351 | B.1.617.2 | B.1.429 | B.1.525 | P.1     | B.1.526 | B.1.617.1 | C37    | P.3   | P.2   |
| S1  | 0.901   | 0       | 0.128     | 98.132  | 0       | 0       | 0.450   | 0         | 0      | 0     | 0.386 |
| S2  | 3.225   | 0       | 0.444     | 91.657  | 0       | 0       | 2.780   | 0         | 0      | 0     | 1.890 |
| S3  | 8.211   | 0.182   | 0.669     | 86.678  | 0.121   | 0       | 3.102   | 0         | 0      | 0     | 1.034 |
| S4  | 5.664   | 0.129   | 0.284     | 89.730  | 0.078   | 0.360   | 2.224   | 0         | 0      | 0     | 1.526 |
| S5  | 5.454   | 0.122   | 0.531     | 90.357  | 0.265   | 0.102   | 2.226   | 0         | 0      | 0     | 0.939 |
| S6  | 9.690   | 0.169   | 0.120     | 84.001  | 0.338   | 0.072   | 3.528   | 0         | 0.024  | 0     | 2.054 |
| S7  | 12.816  | 0.394   | 0.037     | 78.701  | 0.319   | 0.037   | 6.173   | 0         | 0.075  | 0     | 1.444 |
| S8  | 13.509  | 0.305   | 0.051     | 78.666  | 0.368   | 0.089   | 6.042   | 0         | 0.051  | 0     | 0.915 |
| S9  | 20.555  | 0.462   | 0.060     | 68.716  | 0.207   | 0.146   | 7.891   | 0         | 0.036  | 0     | 1.924 |
| S10 | 28.074  | 0.752   | 0.014     | 60.096  | 0.184   | 0.170   | 8.875   | 0         | 0.156  | 0     | 1.675 |
| S11 | 39.221  | 0.848   | 0.066     | 46.540  | 0.613   | 0.379   | 10.767  | 0.011     | 0.100  | 0     | 1.450 |
| S12 | 43.730  | 0.879   | 0.017     | 42.697  | 0.469   | 0.896   | 10.081  | 0.025     | 0.093  | 0.008 | 1.101 |
| S13 | 52.638  | 0.966   | 0.041     | 30.400  | 0.575   | 1.274   | 12.986  | 0.041     | 0.233  | 0.006 | 0.836 |
| S14 | 59.192  | 0.986   | 0.060     | 23.009  | 0.551   | 1.942   | 13.444  | 0.090     | 0.125  | 0.005 | 0.595 |
| S15 | 63.386  | 0.962   | 0.076     | 16.452  | 0.511   | 3.007   | 15.055  | 0.056     | 0.132  | 0.004 | 0.354 |
| S16 | 68.156  | 0.941   | 0.126     | 11.588  | 0.443   | 4.464   | 13.626  | 0.097     | 0.203  | 0.003 | 0.349 |
| S17 | 71.272  | 0.984   | 0.205     | 8.356   | 0.430   | 5.516   | 12.728  | 0.171     | 0.191  | 0.011 | 0.132 |
| S18 | 71.693  | 0.935   | 0.491     | 6.260   | 0.365   | 6.729   | 12.998  | 0.193     | 0.238  | 0     | 0.092 |
| S19 | 72.674  | 0.830   | 1.133     | 4.876   | 0.385   | 7.625   | 11.917  | 0.192     | 0.328  | 0     | 0.037 |
| S20 | 73.363  | 0.914   | 1.560     | 3.653   | 0.350   | 9.314   | 10.406  | 0.117     | 0.280  | 0.003 | 0.031 |
| S21 | 73.655  | 0.907   | 2.258     | 2.952   | 0.285   | 10.116  | 9.386   | 0.086     | 0.325  | 0.011 | 0.014 |
| S22 | 72.514  | 0.733   | 3.193     | 2.273   | 0.176   | 11.428  | 9.279   | 0.038     | 0.343  | 0.014 | 0.004 |
| S23 | 70.802  | 0.851   | 5.351     | 1.491   | 0.157   | 12.237  | 8.661   | 0.018     | 0.422  | 0.006 | 0     |
| S24 | 66.962  | 0.536   | 10.030    | 1.309   | 0.114   | 13.353  | 7.226   | 0.017     | 0.430  | 0.008 | 0.008 |
| S25 | 58.024  | 0.331   | 19.942    | 0.741   | 0.199   | 13.704  | 6.691   | 0.011     | 0.331  | 0     | 0.022 |
| S26 | 49.989  | 0.220   | 32.039    | 0.490   | 0.110   | 12.037  | 4.652   | 0         | 0.460  | 0     | 0     |
| S27 | 39.350  | 0.120   | 46.363    | 0.290   | 0.040   | 10.328  | 3.085   | 0.010     | 0.410  | 0     | 0     |
| S28 | 24.301  | 0.050   | 64.570    | 0.192   | 0.033   | 8.595   | 1.819   | 0         | 0.419  | 0.017 | 0     |
| S29 | 15.025  | 0.062   | 77.618    | 0.062   | 0.013   | 5.887   | 1.099   | 0         | 0.229  | 0     | 0     |
| S30 | 7.749   | 0.028   | 88.004    | 0.028   | 0       | 3.407   | 0.532   | 0         | 0.247  | 0     | 0     |
| S31 | 3.907   | 0.014   | 93.662    | 0.008   | 0.002   | 2.032   | 0.211   | 0         | 0.160  | 0     | 0     |
| S32 | 1.931   | 0.014   | 96.704    | 0.013   | 0       | 1.166   | 0.076   | 0         | 0.093  | 0     | 0     |
| S33 | 1.051   | 0.001   | 98.147    | 0.006   | 0       | 0.697   | 0.043   | 0         | 0.052  | 0     | 0     |
| S34 | 0.480   | 0.002   | 98.651    | 0       | 0       | 0.490   | 0.321   | 0         | 0.053  | 0     | 0     |
| S35 | 0.254   | 0.001   | 99.457    | 0       | 0       | 0.246   | 0.014   | 0         | 0.025  | 0     | 0     |
| S36 | 0.142   | 0.001   | 99.533    | 0.119   | 0       | 0.162   | 0.014   | 0.001     | 0.023  | 0     | 0     |
| S37 | 0.065   | 0.001   | 99.811    | 0.002   | 0       | 0.110   | 0.002   | 0.001     | 0.004  | 0     | 0     |
| S38 | 0.050   | 0.001   | 99.894    | 0.001   | 0       | 0.042   | 0.002   | 0         | 0.006  | 0     | 0     |
| S39 | 0.029   | 0       | 99.935    | 0.001   | 0       | 0.028   | 0.002   | 0         | 0.002  | 0     | 0     |
| S40 | 0.054   | 0       | 99.780    | 0.119   | 0       | 0.039   | 0.005   | 0         | 0      | 0     | 0     |
| S41 | 0.005   | 0       | 99.984    | 0       | 0       | 0.010   | 0       | 0         | 0      | 0     | 0     |
| S42 | 0.008   | 0       | 99.976    | 0       | 0       | 0.015   | 0       | 0         | 0      | 0     | 0     |

**Supplementary Table 2 | The composition (in %) of the *in silico* mixtures of sublineages and their parent lineages**

| | Delta 21 I | Omicron 22 E | Omicron 22 A | Omicron 22 C | Omicron 23 A | Delta 21 A | Omicron 22 D | Delta 21 J | Omicron 21 L | Omicron 22 F | Omicron 21 K | Omicron 22 B |
|---|---|---|---|---|---|---|---|---|---|---|---|---|
| | AY.23 | BQ.1 | BA.4 | BA.2.12.1 | XBB.1.5 | B.1.617.2 | BA.2.75 | AY.27 | BA.2 | XBB | BA.1 | BA.5 |
| s1 | 3.388 | 0 | 0 | 0 | 0 | 0.771 | 0 | 95.813 | 0 | 0 | 0.028 | 0 |
| s2 | 2.722 | 0 | 0 | 0 | 0 | 0.479 | 0 | 96.294 | 0 | 0 | 0.504 | 0 |
| s3 | 1.952 | 0 | 0 | 0 | 0 | 0.332 | 0 | 72.667 | 0 | 0 | 25.049 | 0 |
| s4 | 0.402 | 0 | 0 | 0 | 0 | 0.049 | 0 | 15.134 | 0.025 | 0 | 84.390 | 0 |
| s5 | 0.059 | 0 | 0 | 0.007 | 0 | 0.007 | 0 | 2.830 | 0.197 | 0 | 96.901 | 0 |
| s6 | 0.011 | 0 | 0 | 0.011 | 0 | 0 | 0 | 0.621 | 0.697 | 0 | 98.659 | 0 |
| s7 | 0.010 | 0 | 0 | 0.010 | 0 | 0 | 0 | 0.238 | 1.835 | 0 | 97.907 | 0 |
| s8 | 0 | 0 | 0 | 0.029 | 0 | 0 | 0 | 0.116 | 6.649 | 0 | 93.206 | 0 |
| s9 | 0 | 0 | 0 | 0.419 | 0 | 0 | 0 | 0 | 23.892 | 0 | 75.689 | 0 |
| s10 | 0 | 0 | 0 | 3.291 | 0 | 0 | 0 | 0.084 | 56.793 | 0 | 39.831 | 0 |
| s11 | 0 | 0 | 0 | 12.371 | 0 | 0 | 0 | 0 | 74.570 | 0 | 13.058 | 0 |
| s12 | 0 | 0 | 0.124 | 25.232 | 0 | 0 | 0 | 0 | 70.738 | 0 | 3.844 | 0.062 |
| s13 | 0 | 0 | 0.483 | 41.224 | 0 | 0 | 0 | 0 | 57.005 | 0 | 0.966 | 0.322 |
| s14 | 0 | 0 | 1.701 | 56.207 | 0 | 0 | 0 | 0 | 40.010 | 0 | 0.279 | 1.802 |
| s15 | 0 | 0 | 5.204 | 62.422 | 0 | 0 | 0 | 0 | 24.652 | 0 | 0.120 | 7.602 |
| s16 | 0 | 0 | 10.679 | 51.548 | 0 | 0 | 0 | 0 | 13.487 | 0 | 0.072 | 24.214 |
| s17 | 0 | 0 | 14.433 | 28.500 | 0 | 0 | 0.023 | 0 | 5.641 | 0 | 0.046 | 51.359 |
| s18 | 0 | 0 | 14.010 | 11.685 | 0 | 0 | 0.042 | 0 | 2.179 | 0 | 0.042 | 72.042 |
| s19 | 0 | 0 | 12.354 | 4.526 | 0 | 0 | 0.128 | 0 | 0.967 | 0 | 0.018 | 82.007 |
| s20 | 0 | 0 | 11.735 | 1.797 | 0 | 0 | 0.227 | 0 | 0.409 | 0 | 0.023 | 85.809 |
| s21 | 0 | 0.081 | 11.207 | 0.567 | 0 | 0 | 0.540 | 0 | 0.216 | 0 | 0 | 87.389 |
| s22 | 0 | 0.296 | 12.278 | 0.230 | 0 | 0 | 0.823 | 0 | 0.165 | 0 | 0.033 | 86.175 |
| s23 | 0 | 1.355 | 12.281 | 0.087 | 0 | 0 | 1.267 | 0 | 0.262 | 0.044 | 0 | 84.703 |
| s24 | 0 | 5.076 | 10.773 | 0.056 | 0 | 0 | 1.918 | 0 | 0.508 | 0.282 | 0 | 81.387 |
| s25 | 0 | 12.682 | 9.004 | 0.063 | 0 | 0 | 2.854 | 0 | 0.698 | 0.634 | 0 | 74.065 |
| s26 | 0 | 26.429 | 6.299 | 0 | 0.065 | 0 | 3.571 | 0 | 0.779 | 1.234 | 0.065 | 61.558 |
| s27 | 0 | 42.249 | 3.708 | 0 | 0.304 | 0 | 4.681 | 0 | 0.912 | 2.553 | 0 | 45.593 |
| s28 | 0 | 55.324 | 1.915 | 0 | 1.521 | 0 | 4.901 | 0 | 0.789 | 3.718 | 0 | 31.831 |
| s29 | 0 | 62.393 | 1.084 | 0 | 5.159 | 0 | 4.860 | 0 | 0.598 | 4.075 | 0 | 21.832 |
| s30 | 0 | 63.739 | 0.565 | 0 | 10.780 | 0 | 5.126 | 0 | 0.603 | 3.732 | 0.075 | 15.379 |
| s31 | 0 | 54.398 | 0.333 | 0 | 25.942 | 0 | 4.545 | 0 | 0.443 | 4.028 | 0.037 | 10.273 |
| s32 | 0 | 37.296 | 0.169 | 0 | 50.592 | 0 | 3.268 | 0 | 0.169 | 3.324 | 0 | 5.183 |
| s33 | 0 | 23.976 | 0.060 | 0 | 68.193 | 0 | 2.410 | 0 | 0.181 | 2.229 | 0 | 2.952 |
| s34 | 0 | 13.793 | 0 | 0 | 81.392 | 0 | 1.627 | 0 | 0.130 | 1.627 | 0 | 1.431 |
| s35 | 0 | 8.525 | 0 | 0 | 87.298 | 0 | 1.620 | 0 | 0.085 | 1.705 | 0 | 0.767 |
| s36 | 0 | 5.333 | 0 | 0 | 90.788 | 0 | 1.212 | 0 | 0 | 2.061 | 0 | 0.606 |

**Supplementary Table 3 | The genome references corresponding to the lineages and sublineages in the *in silico* mixtures**

| Pango Lineage Identifier | Alternative Name | GISAID ID |
|---|---|---|
| B.1.1.7 | Alpha | EPI_ISL_17160289 |
| B.1.351 | Beta | EPI_ISL_17138021 |
| B.1.617.2 (Suppl. Table 1) | Delta | EPI_ISL_17226754 |
| B.1.429 | Epsilon | EPI_ISL_17137934 |
| B.1.525 | Eta | EPI_ISL_12828399 |
| P.1 | Gamma | EPI_ISL_17117155 |
| B.1.526 | Iota | EPI_ISL_17160663 |
| B.1.617.1 | Kappa | EPI_ISL_9548730 |
| C37 | Lambda | EPI_ISL_3510626 |
| P.3 | Theta | EPI_ISL_2927545 |
| P.2 | Zeta | EPI_ISL_1391728 |
| AY.23 | Delta 21 I | EPI_ISL_11972471 |
| BQ.1 | Omicron 22 E | EPI_ISL_17253573 |
| BA.4 | Omicron 22 A | EPI_ISL_15112469 |
| BA.2.12.1 | Omicron 22 C | EPI_ISL_17342962 |
| XBB.1.5 | Omicron 23 A | EPI_ISL_17367918 |
| B.1.617.2 (Suppl. Table 2) | Delta 21 A | EPI_ISL_17369462 |
| BA.2.75 | Omicron 22 D | EPI_ISL_16252689 |
| AY.27 | Delta 21 J | EPI_ISL_17138148 |
| BA.2 | Omicron 21 L | EPI_ISL_17159713 |
| XBB | Omicron 22 F | EPI_ISL_16159016 |
| BA.1 | Omicron 21 K | EPI_ISL_13818479 |
| BA.5 | Omicron 22 B | EPI_ISL_17143753 |

**Supplementary Table 4 | The composition (in %) of the *in vitro* mixtures of lineages with proportions similar to those in the corresponding *in silico* mixtures**

| Sample | Alpha B.1.1.7 | Beta B.1.351 | Delta B.1.617.2 | Epsilon B.1.429 | Gamma P.1 | Iota B.1.526 | Kappa B.1.617.1 |
|---|---|---|---|---|---|---|---|
| S1 | 0.998 | 0 | 0.225 | 98.229 | 0 | 0.547 | 0 |
| S2 | 8.443 | 0.414 | 0.901 | 86.910 | 0 | 3.334 | 0 |
| S3 | 71.382 | 1.094 | 0.315 | 8.466 | 5.626 | 12.838 | 0.281 |
| S4 | 73.459 | 1.010 | 1.656 | 3.749 | 9.410 | 10.502 | 0.213 |
| S5 | 72.591 | 0.810 | 3.270 | 2.350 | 11.505 | 9.356 | 0.115 |
| S6 | 50.085 | 0.316 | 32.135 | 0.586 | 12.133 | 4.748 | 0 |
| S7 | 3.935 | 0.042 | 93.690 | 0.036 | 2.060 | 0.239 | 0 |
| S8 | 0.260 | 0.007 | 99.463 | 0 | 0.252 | 0.020 | 0 |
| S9 | 0.067 | 0.003 | 99.813 | 0.002 | 0.112 | 0.002 | 0.003 |
| S10 | 0.008 | 0 | 99.977 | 0 | 0.015 | 0 | 0 |

Mixtures S1 to S10 are adapted from mixtures S1, S3, S17, S20, S22, S26, S31, S35, S37, and S42, respectively from Supplementary Table 1

**Supplementary Table 5 | The composition (in %) of the *in vitro* mixtures of sublineages and their parent lineages with proportions similar to those in the corresponding *in silico* mixtures**

| Sample | Delta AY.1 | Omicron BQ.1 | Omicron BA.4 | Omicron BA.2.12.1 | Delta B.1.617.2 | Delta AY.2 | Omicron BA.2 | Omicron BA.1 | Omicron BA.5 |
|---|---|---|---|---|---|---|---|---|---|
| S1 | 3.388 | 0 | 0 | 0 | 0.771 | 95.813 | 0 | 0.028 | 0 |
| S2 | 0.402 | 0 | 0 | 0 | 0.049 | 15.134 | 0.025 | 84.390 | 0 |
| S3 | 0.010 | 0 | 0 | 0.010 | 0 | 0.238 | 1.835 | 97.907 | 0 |
| S4 | 0 | 0 | 0 | 3.291 | 0 | 0.084 | 56.793 | 39.831 | 0 |
| S5 | 0 | 0 | 0.483 | 41.224 | 0 | 0 | 57.005 | 0.966 | 0.322 |
| S6 | 0 | 0 | 10.679 | 51.548 | 0 | 0 | 13.487 | 0.072 | 24.214 |
| S7 | 0 | 0 | 11.781 | 1.842 | 0 | 0 | 0.455 | 0.068 | 85.854 |
| S8 | 0 | 0.189 | 11.315 | 0.675 | 0 | 0 | 0.324 | 0 | 87.497 |
| S9 | 0 | 1.617 | 12.544 | 0.350 | 0 | 0 | 0.524 | 0 | 84.965 |
| S10 | 0 | 13.380 | 9.702 | 0.761 | 0 | 0 | 1.395 | 0 | 74.762 |

Mixtures S1 to S10 are adapted from mixtures S1, S4, S7, S10, S13, S16, S20, S21, S23, and S25, respectively from Supplementary Table 2

**Supplementary Table 6 | Genome references corresponding to the synthetic SARS-CoV-2 RNA genomes used for the *in vitro* mixtures from Supplementary Tables 4 and 5**

| Pango Lineage Designation | Alternative name | GISAID ID |
|---|---|---|
| B.1.1.7 | Alpha | EPI_ISL_710528 |
| B.1.1.7 | Alpha | EPI_ISL_601443 |
| B.1.351 | Beta | EPI_ISL_678597 |
| P.1 | Gamma | EPI_ISL_792683 |
| B.1.617.1 | Kappa | EPI_ISL_1662307 |
| B.1.526 | Iota | EPI_ISL_1300881 |
| B.1.429 | Epsilon | EPI_ISL_672365 |
| B.1.617.2 | Delta | EPI_ISL_1544014 |
| AY.1 | Delta AY.1 | EPI_ISL_2695467 |
| AY.2 | Delta AY.2 | EPI_ISL_2693246 |
| BA.1 | Omicron BA.1 | EPI_ISL_6841980 |
| BA.2 | Omicron BA.2 | EPI_ISL_7190366 |
| BA.2 | Omicron BA.2 | EPI_ISL_7718520 |
| BA.2.12.1 | - | EPI_ISL_12248637.1 |
| BA.2.12.1 | - | EPI_ISL_12303256.1 |
| BA.4.1 | Omicron BA.4 | EPI_ISL_12516495 |
| BA.5.5 | Omicron BA.5 | EPI_ISL_12620611 |
| BA.4.1 | Omicron BA.4 | EPI_ISL_12454576 |
| BA.4.1 | Omicron BA.4 | EPI_ISL_12605687 |
| BQ.1 | - | EPI_ISL_14829147 |

**Supplementary Table 7 | The composition (in %) of the *in vitro* mixtures with standard lineage and sublineage ratios**

| Sample | original Wuhan-Hu-1 A.1 | alpha B.1.1.7 | beta B.1.351 | gamma P.1 | delta B.1.617.2 | delta AY.1 | delta AY.2 | iota B.1.526 | omicron BA.1 | omicron BA.2 |
|---|---|---|---|---|---|---|---|---|---|---|
| S1 | 25 | 75 | 0 | 0 | 0 | 0 | 0 | 0 | 0 | 0 |
| S2 | 25 | 50 | 25 | 0 | 0 | 0 | 0 | 0 | 0 | 0 |
| S3 | 25 | 25 | 25 | 25 | 0 | 0 | 0 | 0 | 0 | 0 |
| S4 | 50 | 0 | 0 | 0 | 0 | 0 | 0 | 50 | 0 | 0 |
| S5 | 25 | 0 | 0 | 0 | 0 | 0 | 0 | 25 | 0 | 50 |
| S6 | 0 | 25 | 0 | 0 | 0 | 0 | 0 | 25 | 25 | 25 |
| S7 | 0 | 0 | 0 | 0 | 0 | 0 | 0 | 0 | 50 | 50 |
| S8 | 25 | 25 | 0 | 0 | 0 | 0 | 0 | 0 | 25 | 25 |
| S9 | 0 | 0 | 0 | 0 | 0 | 50 | 50 | 0 | 0 | 0 |
| S10 | 0 | 0 | 0 | 0 | 50 | 25 | 25 | 0 | 0 | 0 |
| S11 | 0 | 0 | 0 | 0 | 0 | 25 | 25 | 0 | 50 | 0 |
| S12 | 0 | 0 | 0 | 0 | 0 | 25 | 25 | 0 | 25 | 25 |
| S13 | 0 | 0 | 0 | 0 | 0 | 25 | 25 | 0 | 25 | 25 |
| S14 | 0 | 0 | 0 | 0 | 50 | 0 | 0 | 0 | 25 | 25 |
| S15 | 0 | 0 | 0 | 0 | 0 | 25 | 25 | 0 | 25 | 25 |
| S16 | 0 | 25 | 0 | 0 | 25 | 0 | 0 | 0 | 25 | 25 |

**Supplementary Table 8 | ART tool parameters relevant for our simulation and their interpretation**

| Parameter used | Parameter interpretation (from the ART manual) |
|---|---|
| -i | filename or path of input DNA/RNA reference used as a simulation template |
| -o | prefix of output filename or path |
| -sam | generate SAM alignment file |
| -ss | name of Illumina sequencing system of the built-in profile used for simulation |
| -ef | generate the zero sequencing errors SAM file as well the regular one* |
| -p | paired-end read simulation |
| -l | length of reads to be simulated |
| -c | number of reads/read pairs to be generated per sequence/amplicon |
| -m | mean size of DNA/RNA fragments for paired-end simulations |
| -s | standard deviation of DNA/RNA fragment size for paired-end simulations |
| -ir | first-read insertion rate (default: 0.00009) |
| -ir2 | second-read insertion rate (default: 0.00015) |
| -dr | first-read deletion rate (default: 0.00011) |
| -dr2 | second-read deletion rate (default: 0.00023) |

*reads in the zero-error SAM file have the same alignment positions as those in the regular SAM file (true alignments)